\RequirePackage{fix-cm}
\documentclass[smallextended]{svjour3}       % onecolumn (second format)
\smartqed  % flush right qed marks, e.g. at end of proof
\usepackage{graphicx}
\usepackage{amsmath}
\usepackage[square,sort,comma,numbers]{natbib}
\journalname{Synthesis}
\begin{document}

\newcommand{\tr}{{\rm T}\ensuremath{_{\rm R}}}
\newcommand{\ta}{{\rm T}\ensuremath{_{\rm A}}}
\newcommand{\ts}{{\rm T}\ensuremath{_{\rm S}}}
\newcommand{\vi}{{\rm v}\ensuremath{_{\rm i}}}
\newcommand{\vis}{{\rm v}\ensuremath{_{\rm i}^{*}}}
\newcommand{\vj}{{\rm v}\ensuremath{_{\rm j}}}
\newcommand{\vjs}{{\rm v}\ensuremath{_{\rm j}^{*}}}
\newcommand{\si}{{\rm s}\ensuremath{_{\rm i}}}
\newcommand{\sis}{{\rm s}\ensuremath{_{\rm i}}^{*}}
\newcommand{\ei}{{\rm e}\ensuremath{_{\rm i}}}
\newcommand{\eis}{{\rm e}\ensuremath{_{\rm i}}^{*}}
\newcommand{\sj}{{\rm s}\ensuremath{_{\rm j}}}
\newcommand{\sjs}{{\rm s}\ensuremath{_{\rm j}^{*}}}
\newcommand{\ej}{{\rm e}\ensuremath{_{\rm j}}}
\newcommand{\ejs}{{\rm e}\ensuremath{_{\rm j}}^{*}}
\newcommand{\sk}{{\rm s}\ensuremath{_{\rm k}}}
\newcommand{\sks}{{\rm s}\ensuremath{_{\rm k}^{*}}}
\newcommand{\ek}{{\rm e}\ensuremath{_{\rm k}}}
\newcommand{\eks}{{\rm e}\ensuremath{_{\rm k}}^{*}}
\newcommand{\sls}{{\rm s}\ensuremath{_{\rm l}}^{*}}
\newcommand{\el}{{\rm e}\ensuremath{_{\rm l}}}
\newcommand{\els}{{\rm e}\ensuremath{_{\rm l}}^{*}}
\newcommand{\pwi}{{\rm p}\ensuremath{_{\rm i}}}
\newcommand{\dti}{\ensuremath{\Delta \tau_{i}}}
\newcommand{\dtj}{\ensuremath{\Delta \tau_{j}}}
\newcommand{\dtij}{\ensuremath{\Delta \tau_{ij}}}
\newcommand{\Gp}{\ensuremath{\rm G^{'}}}
\newcommand{\es}{\ensuremath{e^{-\sigma^2}}}
\newcommand{\esh}{\ensuremath{e^{-\sigma^2/2}}}

\title{Phased array observations with infield phasing}
%\subtitle{}

%\titlerunning{Short form of title}        % if too long for running head

\author{Sanjay Kudale         \and
        Jayaram N. Chengalur %etc.
}

%\authorrunning{Short form of author list} % if too long for running head

\institute{S. Kudale \at
              NCRA-TIFR, Pune University Campus, Ganeshkhind, Pune 411007,India \\
              Tel.: +91-20-25719000\\
              Fax: +91-20-25697257\\
              \email{ksanjay@ncra.tifr.res.in}\\
           \and
           Jayaram. N. Chengalur \at
              NCRA-TIFR, Pune University Campus, Ganeshkhind, Pune 411007,India \\
              Tel.: +91-20-25719000\\
              Fax: +91-20-25697257\\
              \email{chengalur@ncra.tifr.res.in}\\
}

\date{Received: date / Accepted: date}
% The correct dates will be entered by the editor

\maketitle

\begin{abstract}

  We present results from pulsar observations using the Giant Metrewave
  Radio Telescope (GMRT) as a phased
  array with infield phasing. The antennas were kept in phase throughout
  the observation by applying antenna based phase corrections derived from
  visibilities  that were obtained in parallel with the phased array beam
  data, and which were flagged and calibrated in real time using a model
  for the continuum emission in the target field. We find that, as expected,
  the signal to noise ratio (SNR) does  not degrade with  time. In contrast
  observations in which the phasing is done only at the start of the
  observation show a clear  degradation of the SNR with time. We find that
  this degradation is well fit by a function of the form
  SNR($\tau$)~=~$\alpha + \beta e^{-(\tau/\tau_0)^{5/3}}$, which corresponds to
  the case where the phase drifts are caused by Kolmogorov type turbulence in
  the ionosphere. We also present general formulae (i.e. including the effects
  of correlated sky noise, imperfect phasing and self noise) for the  SNR
  and synthesized beam size for phased arrays (as well as corresponding
  formulae for incoherent arrays). These would be useful in planning
  observations with large array telescopes.
  
% \PACS{PACS code1 \and PACS code2 \and more}
% \subclass{MSC code1 \and MSC code2 \and more}
\end{abstract}

\section{Introduction}
\label{intro}

High spatial resolution images in radio astronomy are generally made
by the use of interferometric arrays. In situations where high time
resolution is critical and imaging is not that important (for example
in observations of pulsars) the signals from the individual antennas in
the array are usually combined together to produce a single high time
resolution time series \citep{nice92}. For maximum sensitivity the signals
coming from the direction of interest should be added up in phase; in ideal
circumstances for an array of N identical antennas, this would improve
the signal to noise ratio (SNR) by a factor of N \cite{vanleeuwen10}.

At the GMRT this phasing has so far been achieved by interleaving
observations of the target field with observations of a standard calibrator
\citep{bhattacharyya08}. The antenna based gains (which contain both
ionospheric and instrumental contributions) are determined using the data
on the calibrator and then the same are used when observing the target
source. This approach has a number of drawbacks, viz. (1) the ionospheric
contribution to the phase in the direction of the calibrator source could
be different from that in the direction of the target (2) the observations
have to be periodically interrupted in order to phase up the antennas
and (3) since one would like to keep these interuptions to a minimum, the
phasing interval is generally relatively long ($\sim 1$~hour), which could lead
to a non negligible change in the SNR from the start to the end of the
observations and (4) since the phase generally changes fastest on the
long baselines \citep{nice92}, the more distant antennas are
often excluded from the phased array, which reduces the  SNR from the
ideal maximum.

In this paper we describe infield phasing for the GMRT, where
the antenna phases are determined in real time using a model for the
intensity distribution in the target field. This method has none of the
disadvantages listed above.  We present the methodology that has been adopted,
and also show some sample results. We derive general formulae (i.e.
allowing for the possibility of correlated sky noise, imperfect phasing,
non negligible self noise, etc.) for the expected SNR of phased and
incoherent arrays. We also derive an analytical formula for the
the expected degradation of the SNR for a phased array where the antenna
phases drift with time. Finally, we compare the time variation in SNR
observed at the GMRT with that predicted by the analytical formula we derive.
As most of the upcoming (e.g. ASKAP, MEERKAT) and planned (e.g. SKA)
future large radio telescopes are arrays, these forumale, as well as the
technique of infield phasing discussed in this paper would also be of
interest to astronomers planning pulsar observations with these telescopes.

\section{Methodology}
\label{sec:methodology}

The GMRT digital backend can simultaneously produce both the visibilities
in the target field as well as a phased array beam towards a direction
of interest \citep{roy10}. The observed visibilties along with a sky model
for the sources in the target field can be used to generate antenna based
phase solutions (i.e. via ``self-calibration, see e.g. \cite{schwab80,
cornwell81,cornwell89}). The model for the target field can be obtained either
from previous observations (in cases where a field is being observed
repeatedly - for example, for pulsar timing purposes, such observations
are generally available), or from the current observations themselves. In the
absence of baseline based errors and dominant non-isoplanatic effects
(which is a reasonable assumption for the GMRT and for the
  frequencies being considered here), the observed visibilities $V_{ij}$ can
be written as

\begin{equation}
  V_{ij} = g_ig_j^*\widehat{V}_{ij}
  \label{eqn:decomp}
\end{equation}

\noindent where $\hat{V}_{ij}$ are the true visibilities, which are given by 
\begin{equation}
\widehat{V}_{ij}  = \int \int \frac { \widehat{I}(l,m) B(l,m) e^{-2{\pi}i(ul+vm+wn)}} {\sqrt {1-l^2+m^2}} \,dl\,dm
\end{equation}

\noindent where $\widehat{I}(l,m)$ is the true intensity distribution in
the target field, $B(l,m)$ is the antenna primary beam pattern and {$g_i$} are
the antenna based complex gains. In practise if a model is available for the
field, this can be substituted for $\widehat{I}(l,m)$, to get the
model visibilities. In general the set of equations Eqn.~(\ref{eqn:decomp})
form an overdetermined system (since there are many more baselines than
antennas, see e.g. \cite{thompson01} ) and hence the solutions {$g_{i}$}
can be found as those which minimize

\begin {equation}
  L = \sum_{i=1}^{N}\sum_{j=i+1}^{N} w_{ij}|V_{ij} - g_{i}g_{j}^{*}\widehat {V}_{ij} | ^ 2
  \label{eqn:gain}
\end {equation}

\noindent where $w_{ij}$ are suitable weights. In our case these
solutions were obtained using the {\tt flagcal} pipeline \citep{flagcal,
flagcal1}. As described in more detail in these references
{\tt flagcal} is a multi-threaded flagging and calibration pipeline that
first identifies and flags out discrepant data before
determining the gains using an iterative steepest descent based algorithm.
The model used by {\tt flagcal} can be specified either in the form of a
AIPS format clean component ("CC") table, or in the form a CASA style
model, where the model is specified in the form of a pixellated
image. As mentioned above, the assumptions of isoplanaticity as well as
the absence of baseline based errors which is built into eqn.(~\ref{eqn:gain})
generally hold for the GMRT at this observation frequency. At lower
frequencies, the isoplanatic assumption may break down, nonethelss the
antenna based gains that one would determine using infield phasing would
still provide a better approximation to the phase towards the target source
than the phases obtained from some more distant phase calibrator, observed
at a time different from that of the target source observations.

\section{Observations and Results}

We present here results of observations taken for PSR~B0740-28. Two different
observations were made, one with continuous infield phasing and one without. The
details of the observations are given in Table~\ref{tab:obs}. 

\begin{table}
  \begin{tabular}{lll}
    \hline
    Observation Date      &  22 Mar 2016   & 15 Nov 2016\\
    Phasing               &  At Start      & Continous  \\
    Centre Frequency      &  606.7~MHz     & 606.7 MHz  \\
    Bandwidth             &  31.3~MHz      & 31.3~MHz   \\
    No of Channels        &  512           & 256        \\
    No. of antennas used  &  28            & 26         \\
    Non  working antennas &  W06,S06       & C01,E03,W03,S05\\
    \hline
  \end{tabular}
  \caption{Observational Details}
  \label{tab:obs}
\end{table}

\begin{figure}
\includegraphics[angle=270,width=1.0\textwidth]{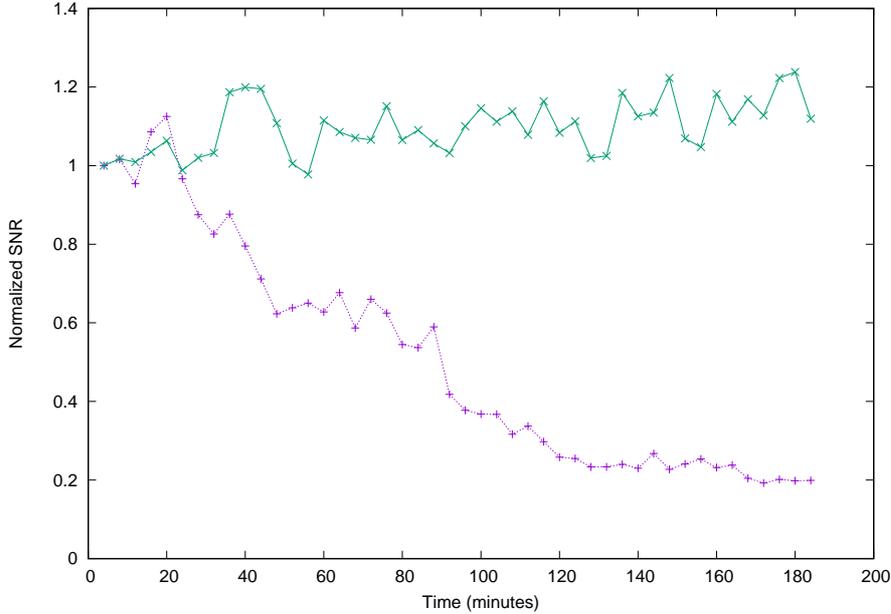}
\caption{ The variation of the normalized SNR with time for the case in which
  the array is phased only at the start of the observations (dotted line)
  as well as the case in which the array is re-phased every 4~min. (solid
  line) using phases derived from the measured visibilities and a model
  of the target field. The normalized SNR is computed over 4~min. intervals and
  normalized by the value of the SNR in the first 4~min.}
\label{fig:snrcmp}     
\end{figure}

For the observations with continuous infield phasing, a model image of the
field was made using data from earlier observations and a
CASA based imaging pipeline (Kudale \& Chengalur, in preparation). Note
that model generation was a one time operation, and the same model was
used for all phasing cycles. The brightest source in the field had a
flux of $\sim 80$~mJy, while the observed flux of the pulsar itself
was $\sim 14$~mJy (these are the observed fluxes, i.e. without correction
for the primary beam. Since the phasing is being done for the same
telescope as the one from which the model was derived, it is the
observed, and not the primary beam corrected flux that is of relevance).
{\tt flagcal} was run on a workstation with a 32 core intel processor
running at 2.6 GHz with 256~GB of RAM. It takes about $\sim 60$~seconds
to flag and calibrate $\sim 3$ minutes of data. We chose to compute and
update the calibration solutions on a time scale of 4~min. The antenna
based phase corrections determined by {\tt flagcal} were uploaded to
the correlator at the end of each 4~min. cycle. {\tt flagcal} provides
both amplitude and phase corrections for each antenna, however only the phases
were updated. This is in keeping with the practise in conventional phasing,
where only the phase derived from the phase calibrator is used to phase
up the array. The phase update also correctly accounts for
the fact that the phases determined by {\tt flagcal} are an incremental
correction over the phases that are currently loaded into the correlator.
For the observations without continuous phasing, the
phases were calculated (using the same model as for the observations
with infield phasing) and applied only at the start of the observations.

The phased array data was de-dispersed using a dispersion measure of
$73.782$~pc~cm$^{-3}$ (i.e. the dispersion measure listed for PSR~B0740-28 
in the ATNF pulsar catalog (http://www.atnf.csiro.au/people/pulsar/psrcat/),
and then each 4~min. time series was folded to obtain the average pulse profile
over that time interval. We show in Fig.~\ref{fig:snrcmp} the normalized
SNR for the two observations. The normalized SNR is the observed SNR in a
given 4~min. interval divided by the SNR for the first 4~min. interval. As can
be seen in the case of infield phasing the SNR is maintained over a period
of several hours, whereas in the absence of phasing the SNR rapidly degrades.
In the next section we derive an general analytical expression for  the
time variation of the degradation of the SNR and compare it with the
observed one.

\subsection{Time variation of the SNR in the absence of continuous
  re-phasing}

Because of drifts in the instrumental phase as well as changes in
the ionospheric conditions, one would expect that in the absence of
re-phasing the SNR of a phased array would gradually decrease with time.
In Appendix~\ref{sec:snr} we derive from first principles the relative SNR
expected for various different kinds of arrays, including phased
arrays in which the phasing is not perfect. We show that under the
assumption that (1) the self noise from the source is negligible and
(2) the phase error on all baselines can be characterized as a Gaussian
random variable with variance $\sigma^2$, the SNR for an N identical
element array scales as

\begin{equation}
  \frac{GS+ (N-1)(GS)\esh}{(\ta + \tr)}
  \label{eqn:snrch}
\end{equation}

\noindent where \tr\ is the ``receiver'' temperature of the antenna (i.e. the
noise contribution from the LNA and any other sources that are independent
for different antennas) and \ta\ is the ``antenna temperature'', i.e.
the noise contribution from the sky and any other sources that are correlated
between the different antennas.

In the case that we are interested in here, viz. that the antennas are
initially phased, but slowly get dephased, the change in the phase with
time can be characterized by the structure function

\begin{equation}
  \sigma^2_{\phi} = <(\phi(t+\tau) - \phi(t))^2>
\end{equation}

\noindent For Kolmogorov turbulence in the ``frozen in'' approximation, the structure
function is given by $D(\tau) \sim \tau^{5/3}$ \citep{thompson01}. The signal
to noise ratio will hence vary as 

\begin{equation}
  {\rm SNR}(\tau) = \frac{(GS) + (N-1)(GS)e^{-(\tau/\tau_0)^{5/3}}}{(\ta + \tr)}
\end{equation}  

\noindent where $\tau_0$ is some constant.

We have assumed above that variance of the phase is the same for all the
baselines in the array. In general however, the phase variation
is faster on the longer baselines \citep{nice92}. For an array like the GMRT
which has about half of the antennas in a compact ``central square'' and
the remainder in sparse extended ``Y'' shaped arms \citep{swarup91}, the
central square antennas generally maintain their phase coherence for
significantly longer than the arm antennas. If we make the simplifying
assumption that over the observing interval some fraction of the antennas
remain phased (i.e. $\sigma^2_{\phi} \approx 0$), while the phase drifts on
the remaining antennas, the degradation in the SNR can then be generically
approximated as
 
\begin{equation}
  {\rm SNR}(\tau) = \alpha + \beta e^{-(\tau/\tau_0)^{5/3}} 
  \label{eqn:snrfit}
\end{equation}

\begin{figure}
  \includegraphics[angle=270,width=1.0\textwidth]{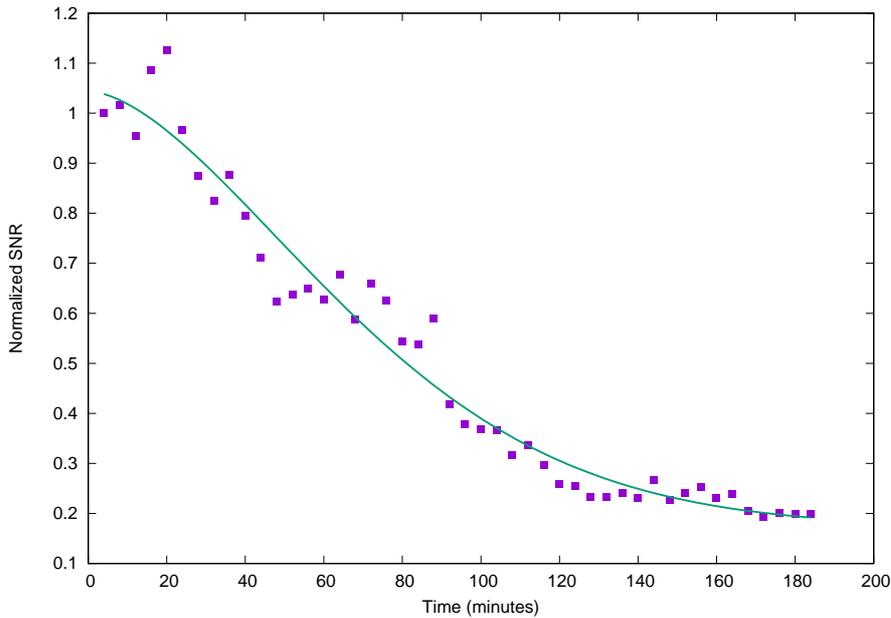}
  \caption{ The variation of the SNR with time (solid points) in the
    case that phasing is  done only at the start of the observation. The
    solid line shows the best fit function of the form
    ${\rm SNR}(\tau) = \alpha + \beta e^{-(\tau/\tau_0)^{5/3}}$. See the text for
    more details. }
\label{fig:snrVsTime}     
\end{figure}

\noindent The functional form given  in Eqn.~(\ref{eqn:snrfit})  was fit to the
data in Fig.~\ref{fig:snrcmp} without continuous infield phasing and
the resulting best
fit along with the original data are shown in Fig.~\ref{fig:snrVsTime}.
As can be seen the expression in Eqn.~(\ref{eqn:snrfit}) gives an excellent fit
to the data. The best fit values of the different parameters are
$\alpha=0.18 \pm 0.02, \beta=0.87 \pm 0.03, \tau_0= 82.1 \pm 4.4$~minutes. In
situations where either conventional phasing is done, or the time interval
between two successive infield phasing cycles is large, Eqn.~(\ref{eqn:snrfit})
allows one to estimate the degradation in SNR in the time interval between
the phase updates. We note that, in general, $\alpha$, $\beta$ and $\tau_0$
would depend on the baseline length distribution in the array as well
as the observing frequency. It would be interesting to try and determine
the frequency dependence of these parameters using future GMRT observations.

\section{Discussion and Conclusions}

We presented a scheme for infield phasing for the GMRT where the
visibility data obtained in parallel with the observations are used
to phase up the voltages used to form the phased array beam.
The primary requirement for such a scheme to work is that there is
sufficient flux in the background continuum sources to allow for
reliable calibration solutions to be found for integration times
of the order of a few minutes. Experience at the GMRT
indicates that this is generally the case in most fields at
frequencies of 610~MHz or lower.

We also present (in the appendix) the expected SNR for different
kinds of arrays, in particular for a phased array in which the
phasing is not perfect. This allows us also to derive a generic
expression for the degradation in the SNR as the antenna phases
drift away from alignment due to ionospheric changes. We note
that the dephasing not only degrades the SNR, but would also
affect the phased array beam; phase errors would have the effect
of broadening the synthesized beam. It is interesting to check
if there is a trade off here -- would introduction of a slight
amount of dephasing (either on purpose, or indavertently) lead
to a faster survey speed? Or does the decrease in signal to noise
ratio out weigh the change in the field of view?

These questions are difficult to answer in general case where the
phase variation is different on different baselines. However,
as discussed above, it is generally the case that the phase errors
increase with baseline length. In order to derive a simple analytical
formula, we make the assumption that the phase errors are such that
$\sigma_{\phi}^2(u,v) \sim 2 a^2 (u^2 +v^2)$ and further that the
baseline density distribution is also Gaussian (i.e. $e^{-b^2(u^2+v^2)}$).
In this case both the degradation in the SNR and the increase in the
synthesized beam size can be analytically computed. In the absence
of phase errors, the synthesized beam (i.e. the image corresponding to
a point source of unit flux at the phase centre) is given by

\begin{equation}
  \int_{-\infty}^{\infty}  \int_{-\infty}^{\infty} dudv\  e^{-b^2(u^2+v^2)} e^{-i2\pi (ul+vm)} \sim e^{-\pi^{2}(l^2+m^2)/b^2}
\end{equation}

Where we have ignorned the normalization, and focussed only on the shape
of the synthesized beam. The area covered at the half maximum level is
clearly $\propto b^2$. In the presense of phase errors, the visbility
on a baseline with separation (u,v) instead of being unity would be
$e^{-\sigma_{\phi}^2(u,v)/2}$ (see Sec.~\ref{sec:snrch}). The synthesized beam
would then be given by

\begin{equation}
  \int_{-\infty}^{\infty}  \int_{-\infty}^{\infty} dudv\ e^{-a^2(u^2+v^2)}
  e^{-b^2(u^2+v^2)} e^{-i2\pi(ul+vm)} \sim e^{-\pi^{2}(l^2+m^2)/(a^2 +b^2)}
\end{equation}

\noindent for which the field of view at the half power level scales as
$(a^2+b^2)$. The field of view on dephasing hence increases by
$\frac{a^2 +b^2}{b^2}$. To get the degradation in the signal to noise ratio we
need to compute the baseline weighted average of $e^{-\sigma_{\phi}^2(u,v)/2}$. This
is given by

\begin{eqnarray}
  <e^{-\sigma_{\phi}^2}> &=& \frac{\int_{-\infty}^{\infty}  \int_{-\infty}^{\infty} dudv\ e^{-(a^2+b^2)(u^2+v^2)}}{ \int_{-\infty}^{\infty}  \int_{-\infty}^{\infty} dudv\ e^{-b^2(u^2+v^2)}}\\
  &=& \frac{b^2}{a^2 + b^2}
\end{eqnarray}

That is, the SNR degrades by the factor $b^2/(a^2 + b^2)$,
which means that one would have to integrate longer by the square of
this factor in order to reach the same SNR as that of a perfectly
phased array. The ratio of survey speeds for the perfectly phased array
to partially dephased array is hence $b^2/(a^2 + b^2)$, i.e.
the partially dephased array has a slower survey speed than the perfectly
phased one.

To summarize, we present data from pulsar observations done at the GMRT using
infield phasing with phase corrections derived from the visibilities obtained
towards the target field, and updated every 4~min. We find that, as expected,
with this quasi-continuous phasing the sensitivity of the phased array can be
sustained over long periods of time. In contrast, the sensitivity in the
situation where the array is phased only at the start of the observations falls
rapidly. We find that the drop in sensitivity is consistent with that expected
for phase fluctuations caused by Kolmogorov type turbulence in the ionosphere.
Infield phasing has the advantage of not only maintaining the sensitivity
of the array, but also of (1) improving the observing efficiency (since one does
not have to periodically slew to a calibrator source) and also, (2) for the
same reason enabling long continuous data runs on a given target.

\appendix

\section{APPENDIX: Array Signal to Noise Ratios }
\label{sec:snr}
We derive here from first principles the degradation in the signal to noise
ratio (SNR) for an imperfectly phased array. For completeness we present
the derivation also for an incoherent array. The derivation has (as would
be expected) similarities with the SNR for imaging arrays \citep{thompson01,
rohlfs96} but we briefly summarize all the steps in the derivation, so
that this presentation stands by itself.

Consider an array of N identical antennas, with ``receiver'' temperature \tr\,
(defined more precisely below) and gain G (in units of K/Jy). That is an
unresolved 1~Jy source at the centre of the primary beam produces an
antenna temperature of G~K. Let us assume that all of the antennas are
observing a source of flux S that is located at the centre of primary
beam of all the antennas which is also the phase centre of the observations.
We further assume that the sky is empty apart from the source at the phase
center. (We relax these requirements below). Since our primary aim is to
compare the SNR in different scenarios , we ignore the effect of other
parameters such as the bandwidth, integration time, number of
polarisations, etc. which we will assume to be
kept the same for all the situations that we discuss.

The voltage signal \vi\ from an individual element is given by
$ \vi = \si + \ei$,  where \si\ is the signal voltage and \ei\ is the noise
voltage. We will assume that both of these have a zero mean Gaussian
distribution. So $<\si> = <\ei> = 0$. Further since the
signal and noise, as well as the noise from different antennas are independent,
we have $<\si \ei> = 0$, and $<\ei \ej> = 0$ for $i \neq j$. Here we are
explicitly assuming that \ei\ contains no component that is correlated
between antennas, i.e. arises entirely from the receiver noise, including
ground pickup etc.(We discuss below the modifications that arise when
these assumptions are relaxed). Under the assumptions listed
above, we have $<\si\sjs> = {\rm GS}$ and $<\ei \eis> = \tr$.

\subsection{Incoherent Arrays}
\label{sec:snrich}

The power \pwi\ measured by antenna $i$ is given by $\pwi = \vi^2$. If the
power from the incoherent array is $y$, we have

$$ y = \sum\limits_{i=1}^N \pwi $$

We hence have

\begin{equation}
  <y> = \sum\limits_{i} <\si^2 + \eis \si + \sis \ei + \ei^2> = {\rm N(GS + \tr)}
\label{eqn:yich}  
\end{equation}

The ``signal'' part of this is NGS. In order to compute the ``noise'' part
we need to compute the variance of $y$. As a first step we compute $<y^2>$.
We have

\begin{eqnarray}
  <y^2> &=& < \sum\limits_{i} (\si + \ei)^2 \sum\limits_{j}(\sj + \ej)^2 > \\
  &=& <\sum\limits_{i}\sum\limits_{j} (\si + \ei)(\sis + \eis)
  (\sj + \ej)(\sjs + \ejs) >
  \label{eqn:y2ichA}
\end{eqnarray}

For a Gaussian distribution, the fourth moment can be written in terms of
products of the second moment, (see e.g. \cite{shynk12}) viz.
\begin{equation}
  \begin{split}
  <(\si + \ei)(\sis + \eis)(\sj + \ej)(\sjs + \ejs) > &=
  <(\si + \ei)(\sis + \eis)><(\sj + \ej)(\sjs + \ejs) > \\
  &+ <(\si + \ei)(\sj + \ej)><(\sis + \eis)(\sjs + \ejs) > \\
  &+ <(\si + \ei)(\sjs + \ejs)><(\sis + \eis)(\sj + \ej) >
  \end{split}
\label{eqn:y2ichB}
\end{equation}

The second term in the sum is zero\footnote{Since for our zero mean 
complex random variable we have $z = x+iy$, $<zz> = <x^2> - <y^2> + 2i<xy>$,
and in the situation considered here we  have $<x^2> = <y^2>$  and
$<xy> = 0$. Hence only terms which involve a variable and its complex
conjugate survive the averaging process}. To evaluate the remaing two terms, let us consider two
separate cases, one when $i = j$, and the other when $i \neq j$.
When $i = j$, the sum evaluates to 2(GS +\tr)$^2$, and there are N such terms.
A term with $i \neq j$ evaluates to $(GS)^2 + (GS+\tr)^2$ (where we have
used the fact that $\si = \sj$, i.e. the sky signal is 100\% correlated).
There are N(N-1) terms with $i \neq j$. Putting all of this together we get

\begin{equation}
  <y^2>= 2N({\rm GS + \tr})^2 + N(N-1)[({\rm GS + \tr})^2 + ({\rm GS})^2]
  \label{eqn:y2ichC}
\end{equation}

\noindent and hence
\begin{equation}
  \sigma_{y}^{2} = <y^2> - <y>^2 = N[({\rm GS + \tr})^2 + (N-1)(GS)^2]
  \label{eqn:sigich}
\end{equation}

\noindent The signal to noise ratio for the incoherent array is hence
\begin{equation}
  {\rm SNR} = \frac{\sqrt{N} {\rm GS}}{[({\rm GS+\tr})^2
      + (N-1)({\rm GS})^2]^{1/2}}
\label{eqn:snrich}        
\end{equation}

Let us now relax the condition that there is only one source in the sky.
To understand the situation where we have multiple sources (or wide spread
diffuse emission) it is useful to first consider the situation where we
still have only one source of flux S in the sky, except that it is not at
the phase and pointing center. For the $i^{th}$ antenna let \dti\ be the
differential geometric delay between the phase center and the position
of the source. It is assumed that all delay measurements are with respect
to some reference position in the array, and the delays can hence be
treated as being antenna based. If \si\ is the signal voltage at the
$i^{th}$ antenna, we then have $<\si \sis> = \Gp S$, where \Gp\ is the
gain in the direction of the source. Further
\begin{equation}
  <\si \sjs> = \Gp S e^{-i2\pi\nu\dtij}
  \label{eqn:visphase}
\end{equation}

\noindent where $\dtij = \dti - \dtj$. Clearly, $<\sj \sis> = \Gp S e^{i2\pi\nu\dtij}$.
Using these relations to compute $<y>$ and $<y^2>$ as above, we find
that the non zero phase terms cancel out and the only change is that G is to be
replaced by \Gp. This is again understandable, since we are dealing with
the power from each antenna, and the signal phase clearly cannot make a
difference to the total power. One assumption that we have made is
that the bandwidth and the integration time are such that time and
bandwidth smearing effects can be ignored, which is a reasonable
assumption for sources within the main beam of the individual antennas.
For a collection of point sources, the power from each source is additive,
so if we have multiple point sources in the sky, Eqn.~(\ref{eqn:snrich})
continues to hold, except that the term GS in the denominator has to be
replaced by \ta, the antenna temperature. Similarly diffuse emission can
be regarded as emission from a collection of point sources distributed
uniformly across the sky, and so once again Eqn.~(\ref{eqn:snrich}) applies
with GS replaced by \ta. In general therefore, the signal to noise
ratio of an incoherent array becomes
\begin{equation}
  {\rm SNR} = \frac{\sqrt{N} {\rm GS}}{[({\rm \ta+\tr})^2
      + (N-1)\ta^2]^{1/2}}
\label{eqn:snrichA}        
\end{equation}

\begin{figure}
  \begin{centering}
    \includegraphics[angle=0,width=0.75\textwidth]{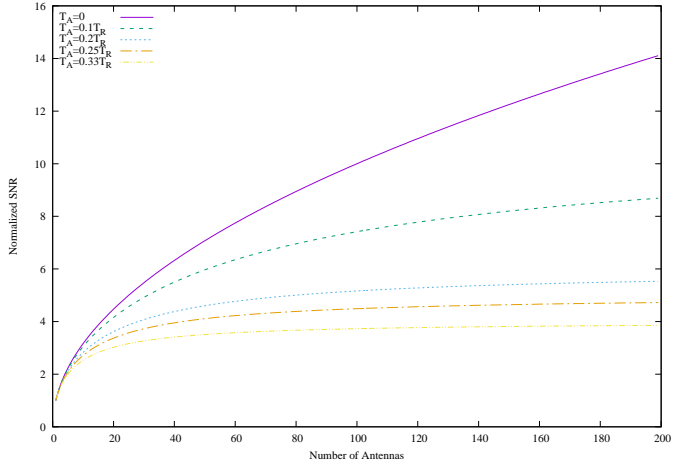}
  \end{centering}
\caption{The variation of the SNR of an incoherent array as a function of
  the number of elements $N$. The solid line shows the ideal case where the
  SNR increases as $\sqrt{N}$ (which requires $\ta =0$),
  the thick dashed line is for $\ta = 0.1\tr$ (which roughly corresponds
  to the GMRT 610~MHz band), the thin dashed line for $\ta = 0.2\tr$, the
  bold dashed dot line for $\ta = 0.25\tr$, and the thin dashed dot line
  is for $\ta = 0.33\tr$ (which roughly corresponds to the expected value for
  SKA-mid \citep{dewdney13}. As discussed in the text, for arrays with
  large number of elements, a ``fly's eye'' mode of observation with
  sub-arrays would be a more efficient survey mode than using all of the
  antennas in the array to form a single incoherent array.}
\label{fig:snrich}     
\end{figure}

Clearly in situations where the sky noise is negligible the SNR will improve
as $\sqrt{N}$; in the opposite extreme when the total noise is sky dominated,
the SNR is independent of $N$ for large $N$. An plot of the increase in the SNR
with N for different values of $\ta/\tr$ is shown in Fig.~\ref{fig:snrich}.
As can be seen in situations where \ta\ is a reasonable fraction of \tr\
one quickly reaches a stage where the SNR saturates and there is diminishing
returns from adding further antennas into the array. In such situations,
surveys would benefit from using a ``fly's eye'' mode of observation with
sub-arrays (i.e. where different sub-arrays look at different regions of
the sky) instead of adding further antennas into the array.

\subsection{Phased Arrays}
\label{sec:snrch}

Let us again start with the scenario where the array is observing an unresolved
source of flux S which is at the phase and pointing center. Let us further
assume that the antenna temperature (due to all the sources including
diffuse emission) is \ta, but that this emission is completely resolved
out on all baselines (i.e. when summed over all the sources in the field
of view the phase term in Eqn.~(\ref{eqn:visphase}) averages out to zero).
Then the only correlation between the voltages at different antennas is
that arising from the source at the phase centre.

Let $ \vi = \si + \ei$ be the voltage from the $i^{th}$ antenna. We 
assume that the phase difference $\phi_{ij}$ between the signals from the
$i^{th}$ and $j^{th}$ antenna has a Gaussian distribution with zero mean
and variance $\sigma^2$. Then $<\vi \vjs> = GS<e^{i\phi_{ij}}> = GS\esh$,
($i \neq j$), and $<\vi \vis> = \ta +\tr$. For simplicity we assumed that
the phase variance $\sigma^2$ is the same for all baselines. Let the total
power output of the phased array be $y$, i.e.
$y = (\sum\limits_{i} \si + \ei)^2$. The expected value of $y$ is then
given by

\begin{eqnarray}
  <y> &=& <\sum\limits_{i}(\si +\ei) \sum\limits_{j}(\sjs +\ejs)>\\
       &=& \sum\limits_{i,j} <\si \sjs> + <\ei \ejs>\\
       &=& N(\ta+\tr) + N(N-1)GS\esh\\
  \label{eqn:ych}
\end{eqnarray}

where we have used the fact that in the summation there are N terms
with $i=j$ (i.e. auto-correlations), and N(N-1) terms with $i \neq j$
(i.e. cross-correlations, which are also called ``visibilities'' in
the imaging context).

As before, in order to estimate the variance of $y$ let us first compute
$<y^2>$. We have
\begin{equation}
  <y^2> = <\sum\limits_{i}(\si +\ei) \sum\limits_{j}(\sjs +\ejs)
           \sum\limits_{k}(\sk +\ek) \sum\limits_{l}(\sls +\els)>
\end{equation}

\noindent Again, as before this $4^{th}$ moment can be written as a product of second moments
and evaluated. We omit the calculation and only give the final result,
which is

\begin{equation}
  <y^2> = 2N^2(\ta+\tr)^2 + 4N^2(N-1)(GS)(\ta+\tr)\esh + 2N^2(N-1)^2(GS)^2\es
  \label{eqn:y2ch}
\end{equation}

\noindent From Eqns.~(\ref{eqn:y2ch})~and~(\ref{eqn:ych}) we get the variance
of $y$ to be
\begin{equation}
  \sigma^2_y = N^2(\ta+\tr)^2 + 2N^2(N-1)(GS)(\ta+\tr)\esh + N^2(N-1)^2(GS)^2\es
  \label{eqn:sigch}  
\end{equation}

\noindent and hence the signal to noise ratio is
\begin{equation}
  \frac{GS +(N-1)(GS)\esh}{[(\ta+\tr)^2+(N-1)(GS)(\ta+\tr)\esh
      +(N-1)^2(GS)^2\es ]^{1/2}}
  \label{eqn:snrchA}
\end{equation}

It is worth high lighting a couple of limiting cases of Eqn.~(\ref{eqn:snrchA}).
In the case where the phasing is perfect (i.e. $\sigma\to 0$) we have
\begin{equation}
  {\rm SNR} = \frac{N({\rm GS})}{[(\ta+\tr)^2+(N-1)(GS)(\ta+\tr)
      +(N-1)^2(GS)^2]^{1/2}}
  \label{eqn:snrchB}
\end{equation}

\noindent which reduces to N(GS)/(\ta+\tr) in the case that ${\rm GS} << (\ta+\tr)$.
Note that this expression holds even if $\ta >> \tr$. In the case that
$\ta << \tr$, the SNR of the phased array  is $\sqrt(N)$ times better
than that of an incoherent array. In the other extreme (i.e. when
($\ta >> \tr$) and for large $N$, the SNR of the phased array is $N$
times better than that of the incoherent array. On the other hand when
the phase is completely random, (i.e. is uniformly distributed over
[$0,2\pi$]), then $<\si,\sjs>$ =0, then the signal to noise ratio
reduces to (GS)/(\ta+\tr) = GS/\ts), i.e. the same as for a single
dish. Essentially, in this case, both the signal
as well as the noise increase identically (since the signal is
not phased up) when one combines together the voltages from different
antennas. In such a situation, the SNR of a ``phased'' array would
in fact be worse than that of an incoherent array.

\begin{figure}
  \begin{centering}
    \includegraphics[angle=0,width=0.75\textwidth]{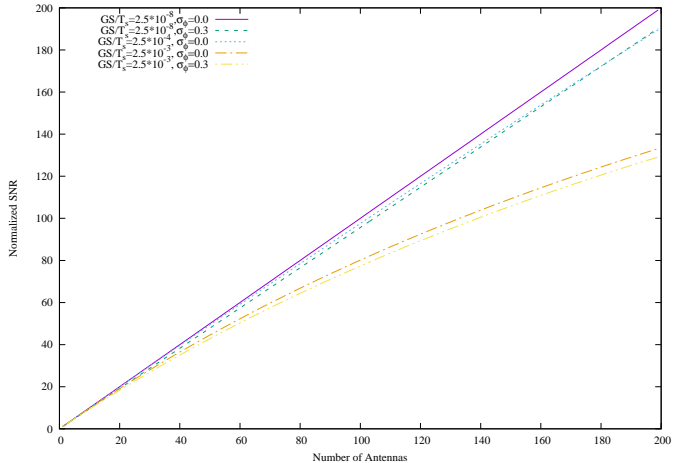}
  \end{centering}
  \caption{The variation of the normalized SNR for a phased array as a
    function of the number of elements $N$. The normalized SNR is the SNR
    divided by the SNR for a single antenna. The solid line shows the ideal
    case where the SNR increases as $N$ (which requires the source flux to be
    negligible, and the phase error to be $0$),
    the thick dashed line is for a negligible source flux and phase rms
    of $0.3$~radians,
    the thin dashed line for $GS/\ts = 2.5\times 10^{-4}$ (which roughly
    corresponds to observation of a 100~mJy source with the SKA-mid,
    \citep{dewdney13})and a phase rms of $0.0$,
    the bold dashed dot line for $GS/\ts = 2.5\times 10^{-3}$  and a phase rms
    of $0.0$,
    and the thin dashed dot line is for $GS/\ts = 2.5\times 10^{-3}$
    and a phase rms of $0.3$~rad. As discussed in the text in the limiting case
    where the antenna phases are uniformly distributed over [$0,2\pi$], the SNR
    does not increase with increasing the number of antennas in the array.}
\label{fig:snrch}     
\end{figure}

\subsection{Post Correlation Beam Formation}
\label{sec:pc}

When forming the coherent phased array beam we assumed that the antenna
voltages were added together and then squared, in order to get the phased
array beam. In an interferometric array one could also produce the
phased array beam from the visibilities produced by the correlator,
a technique sometimes called ``post-correlation beamforming''. If
we exclude the N auto-correlations, then, from Eqn.~(\ref{eqn:ych})
the mean value of the power in the post-correlation beam is
$N(N-1)GS\esh$. Although this is smaller than the corresponding
number for the phased array beam, it does not contain terms proportional
to the receiver temperature \tr\, and is hence likely to be less subject
to systematic variations produced by fluctuations in the receiver power,
ground pick up etc. Following similar arguments as given above for
the phased array, the signal to noise ratio for post-correlation
beam can be shown to be

\begin{equation}
  \frac{(\sqrt{N(N-1)}(GS)\esh}{[(\ta+\tr)^2
      +2(N-2)(GS)(\ta+\tr)\esh + ((N-1)(N-2)+1)(GS)^2\es]^{1/2}}
\end{equation}

\noindent which in the limit of small $S$ is $\sqrt{N(N-1)}/N$ times smaller
than that of the conventional phased array. This is a small factor for
large N.

\subsection{Incoherent Combination of Phased Arrays}
\label{sec:inchpa}

We could also consider the situation (which occurs for example
in the LOFAR array, \cite
{haarlem13}) where we have two levels of array
formation. First
an coherent array is formed from a cluster of nearby elements (a
``station'') and then an incoherent array is formed from these stations.
If there are N elements (each of gain G) in a station, and M such stations,
then the expected value of the total power is $MN(\ta+\tr)+ MN^2GS$. We
have assumed that (1) all the elements in a station are perfectly phased,
and (2) that the diffuse sky emission is dominant and (3) that the power
in the cross-correlation between the elements in a station is small
compared to the power in the auto-correlations (i.e. the terms involving
\es\ and \esh\ in eqn.(~\ref{eqn:snrchA}) can be ignored). Essentially
we are assuming that the diffuse emission is mostly resolved out on the
baselines between the elements in a station.  We note that this is
an extreme assumption, and that in real life arrays, while the
cross-correlation   power is less than the auto-correlation power, it
may not be so small as to be neglected altogether.

For the variance in the final signal formed by incoherently combining
the powers of the stations we will have to compute
\begin{equation}
<y^2>=  <\sum\limits_{m=1}^{M}\sum\limits_{n=1}^{M}\sum\limits_{j=1}^{N}\sum\limits_{j=1}^{N} \sum\limits_{k=1}^{N}\sum\limits_{l=1}^{N}<v_{mi}v_{mj}^{*}v_{nk}v_{nl}^{*}>
\end{equation}

Since we have assumed that diffuse emission dominates, and that it is resolved
out on the baselines between the elements in a station, it follows that it
would also be resolved out on the baselines between the elements in different
stations. So in the summation above, one set of terms which contribute are
those with $n=m$, i.e. the elements within a given station. This contribution
is given by $2N^2(\ta + \tr)^2$. There will be $M$ such
terms. For $n \neq m$, the only terms in the $4^{th}$ moment that will
contribute are of the form $ <v_{mi}v_{mi}^{*}><v_{nk}v_{nk}^{*}>$. There
are $M(M-1)N^2$ such terms, each of which evaluates to  $(\ta + \tr)^2$.
We hence have

\begin{eqnarray}
  \sigma_y^2 &=& M(M-1)N^2(\ta + \tr)^2 +2MN^2(\ta + \tr)^2 -M^2N^2(\ta+\tr)^2\\
        &=& MN^2(\ta + \tr)^2\\
\end{eqnarray}

\noindent The signal to noise ratio is hence
\begin{equation}
  {\rm SNR} = \frac{\sqrt{M}N(GS)}{(\ta + \tr)}
\end{equation}  

\noindent that is, it increases as $\sqrt{M}$ (i.e. the square root of the
number of stations), even though the noise on the
individual elements is sky dominated (i.e. $\ta > \tr$).

\begin{acknowledgements}

  We are grateful to the staff of the GMRT who made these observations
  possible. Helpful discussions with Rajaram Nityananda, Somnath
  Bharadwaj and Jayanta Roy are also gratefully acknowledged. We are
  also grateful for very useful comments on the original version of this
  paper from B. Bhattacharya, and the two anonymous referees.
  
\end{acknowledgements}

\bibliographystyle{spphys}       % APS-like style for physics
\bibliography{infield}   % name your BibTeX data base

\begin{thebibliography}{10}
\providecommand{\url}[1]{{#1}}
\providecommand{\urlprefix}{URL }
\expandafter\ifx\csname urlstyle\endcsname\relax
  \providecommand{\doi}[1]{DOI \discretionary{}{}{}#1}\else
  \providecommand{\doi}{DOI \discretionary{}{}{}\begingroup
  \urlstyle{rm}\Url}\fi

\bibitem{nice92}
D.J. {Nice}, S.E. {Thorsett}, ApJ \textbf{397}, 249 (1992).
\newblock \doi{10.1086/171784}

\bibitem{vanleeuwen10}
J.~{van Leeuwen}, B.W. {Stappers}, A\&A \textbf{509}, A7 (2010).
\newblock \doi{10.1051/0004-6361/200913121}

\bibitem{bhattacharyya08}
B.~{Bhattacharyya}, Y.~{Gupta}, J.~{Gil}, MNRAS \textbf{383}, 1538 (2008).
\newblock \doi{10.1111/j.1365-2966.2007.12666.x}

\bibitem{roy10}
J.~{Roy}, Y.~{Gupta}, U.L. {Pen}, J.B. {Peterson}, S.~{Kudale}, J.~{Kodilkar},
  Experimental Astronomy \textbf{28}, 25 (2010).
\newblock \doi{10.1007/s10686-010-9187-0}

\bibitem{schwab80}
F.R. {Schwab}, in \emph{1980 International Optical Computing Conference I},
  \emph{Proc. of the SPIE}, vol. 231, ed. by W.T. {Rhodes} (1980), \emph{Proc.
  of the SPIE}, vol. 231, pp. 18--25.
\newblock \doi{10.1117/12.958828}

\bibitem{cornwell81}
T.J. {Cornwell}, P.N. {Wilkinson}, MNRAS \textbf{196}, 1067 (1981).
\newblock \doi{10.1093/mnras/196.4.1067}

\bibitem{cornwell89}
T.~{Cornwell}, E.B. {Fomalont}, in \emph{Synthesis Imaging in Radio Astronomy},
  \emph{Astronomical Society of the Pacific Conference Series}, vol.~6, ed. by
  R.A. {Perley}, F.R. {Schwab}, A.H. {Bridle} (1989), \emph{Astronomical
  Society of the Pacific Conference Series}, vol.~6, p. 185

\bibitem{thompson01}
A.R. Thompson, J.M. Moran, G.W.J. Swenson, \emph{Interferometry and Synthesis
  in Radio Astronomy}, 2nd edn. (Wiley Interscience, 2001)

\bibitem{flagcal}
J.~{Prasad}, J.~{Chengalur}, Experimental Astronomy \textbf{33}, 157 (2012).
\newblock \doi{10.1007/s10686-011-9279-5}

\bibitem{flagcal1}
J.N. {Chengalur}, {FLAGCAL: a flagging and calibration pipeline for GMRT DATA}.
\newblock Tech. Rep. NCRA/COM/001, NCRA-TIFR (2013)

\bibitem{swarup91}
G.~{Swarup}, S.~{Ananthakrishnan}, V.K. {Kapahi}, A.P. {Rao}, C.R.
  {Subrahmanya}, V.K. {Kulkarni}, Current Science, Vol.~60, NO.2/JAN25, P.~95,
  1991 \textbf{60}, 95 (1991)

\bibitem{rohlfs96}
K.~Rohlfs, T.L. Wilson, \emph{Tools of Radio Astronomy}, 2nd edn. (Springer,
  1996)

\bibitem{shynk12}
J.Y. Shynk, \emph{Probability, Random Variables and Random Processes: Theory
  and Signal Processing Applications} (Wiley-Interscience, 2012)

\bibitem{dewdney13}
P.E. {Dewdney}, {SKA1 System Baseline Design}.
\newblock Tech. Rep. SKA-TEL-SKO-DD-001, SKA Office (2013)

\bibitem{haarlem13}
M.P. {van Haarlem}, M.W. {Wise}, A.W. {Gunst}, G.~{Heald}, J.P. {McKean},
  J.W.T. {Hessels}, A.G. {de Bruyn}, R.~{Nijboer}, J.~{Swinbank}, R.~{Fallows},
  M.~{Brentjens}, A.~{Nelles}, R.~{Beck}, H.~{Falcke}, R.~{Fender},
  J.~{H{\"o}randel}, L.V.E. {Koopmans}, G.~{Mann}, G.~{Miley},
  H.~{R{\"o}ttgering}, B.W. {Stappers}, R.A.M.J. {Wijers}, S.~{Zaroubi},
  M.~{van den Akker}, A.~{Alexov}, J.~{Anderson}, K.~{Anderson}, A.~{van
  Ardenne}, M.~{Arts}, A.~{Asgekar}, I.M. {Avruch}, F.~{Batejat},
  L.~{B{\"a}hren}, M.E. {Bell}, M.R. {Bell}, I.~{van Bemmel}, P.~{Bennema},
  M.J. {Bentum}, G.~{Bernardi}, P.~{Best}, L.~{B{\^i}rzan}, A.~{Bonafede}, A.J.
  {Boonstra}, R.~{Braun}, J.~{Bregman}, F.~{Breitling}, R.H. {van de Brink},
  J.~{Broderick}, P.C. {Broekema}, W.N. {Brouw}, M.~{Br{\"u}ggen}, H.R.
  {Butcher}, W.~{van Cappellen}, B.~{Ciardi}, T.~{Coenen}, J.~{Conway},
  A.~{Coolen}, A.~{Corstanje}, S.~{Damstra}, O.~{Davies}, A.T. {Deller}, R.J.
  {Dettmar}, G.~{van Diepen}, K.~{Dijkstra}, P.~{Donker}, A.~{Doorduin},
  J.~{Dromer}, M.~{Drost}, A.~{van Duin}, J.~{Eisl{\"o}ffel}, J.~{van Enst},
  C.~{Ferrari}, W.~{Frieswijk}, H.~{Gankema}, M.A. {Garrett}, F.~{de Gasperin},
  M.~{Gerbers}, E.~{de Geus}, J.M. {Grie{\ss}meier}, T.~{Grit}, P.~{Gruppen},
  J.P. {Hamaker}, T.~{Hassall}, M.~{Hoeft}, H.A. {Holties}, A.~{Horneffer},
  A.~{van der Horst}, A.~{van Houwelingen}, A.~{Huijgen}, M.~{Iacobelli},
  H.~{Intema}, N.~{Jackson}, V.~{Jelic}, A.~{de Jong}, E.~{Juette}, D.~{Kant},
  A.~{Karastergiou}, A.~{Koers}, H.~{Kollen}, V.I. {Kondratiev}, E.~{Kooistra},
  Y.~{Koopman}, A.~{Koster}, M.~{Kuniyoshi}, M.~{Kramer}, G.~{Kuper},
  P.~{Lambropoulos}, C.~{Law}, J.~{van Leeuwen}, J.~{Lemaitre}, M.~{Loose},
  P.~{Maat}, G.~{Macario}, S.~{Markoff}, J.~{Masters}, R.A. {McFadden},
  D.~{McKay-Bukowski}, H.~{Meijering}, H.~{Meulman}, M.~{Mevius},
  E.~{Middelberg}, R.~{Millenaar}, J.C.A. {Miller-Jones}, R.N. {Mohan}, J.D.
  {Mol}, J.~{Morawietz}, R.~{Morganti}, D.D. {Mulcahy}, E.~{Mulder}, H.~{Munk},
  L.~{Nieuwenhuis}, R.~{van Nieuwpoort}, J.E. {Noordam}, M.~{Norden},
  A.~{Noutsos}, A.R. {Offringa}, H.~{Olofsson}, A.~{Omar}, E.~{Orr{\'u}},
  R.~{Overeem}, H.~{Paas}, M.~{Pandey-Pommier}, V.N. {Pandey}, R.~{Pizzo},
  A.~{Polatidis}, D.~{Rafferty}, S.~{Rawlings}, W.~{Reich}, J.P. {de Reijer},
  J.~{Reitsma}, G.A. {Renting}, P.~{Riemers}, E.~{Rol}, J.W. {Romein},
  J.~{Roosjen}, M.~{Ruiter}, A.~{Scaife}, K.~{van der Schaaf}, B.~{Scheers},
  P.~{Schellart}, A.~{Schoenmakers}, G.~{Schoonderbeek}, M.~{Serylak},
  A.~{Shulevski}, J.~{Sluman}, O.~{Smirnov}, C.~{Sobey}, H.~{Spreeuw},
  M.~{Steinmetz}, C.G.M. {Sterks}, H.J. {Stiepel}, K.~{Stuurwold}, M.~{Tagger},
  Y.~{Tang}, C.~{Tasse}, I.~{Thomas}, S.~{Thoudam}, M.C. {Toribio}, B.~{van der
  Tol}, O.~{Usov}, M.~{van Veelen}, A.J. {van der Veen}, S.~{ter Veen}, J.P.W.
  {Verbiest}, R.~{Vermeulen}, N.~{Vermaas}, C.~{Vocks}, C.~{Vogt}, M.~{de Vos},
  E.~{van der Wal}, R.~{van Weeren}, H.~{Weggemans}, P.~{Weltevrede},
  S.~{White}, S.J. {Wijnholds}, T.~{Wilhelmsson}, O.~{Wucknitz},
  S.~{Yatawatta}, P.~{Zarka}, A.~{Zensus}, J.~{van Zwieten}, A\&A \textbf{556},
  A2 (2013).
\newblock \doi{10.1051/0004-6361/201220873}

\end{thebibliography}

\end{document}